\documentstyle[12pt]{article}

\begin{document}
\def\be{\begin{equation}}
\def\bea{\begin{eqnarray}}
\def\eea{\end{eqnarray}}
\def\ee{\end{equation}}
\def\Chr#1#2{\left\{^{#1}_{#2}\right\}}
\def\half{\frac{1}{2}}
\def\rg{\sqrt{-g}}
\def\A{{\cal A}}
\def\a{\alpha}
\def\t{\tau}
\def\b{\beta}
\def\m{\mu}
\def\n{\nu}
\def\k{\kappa}
\def\g{\gamma}
\def\G{\Gamma}
\def\e{\eta}
\def\E{\varepsilon}
\def\l{\lambda}
\def\L{{\cal L}}
\def\Rt{\tild{R}}
\def\Ne{$\nu_e\;$}
\def\Nm{$\nu_\mu\;$}
\def\c{\cite}
\def\la{\label}
\def\r{\ref}
\begin{titlepage}
\vspace*{10mm}
\begin{center} {\Large \bf Quantum reflection of massless neutrinos \\
\vskip 1cm
from a torsion--induced potential} \\
\vskip 10mm
\centerline {\bf
M. Alimohammadi$^a$ \footnote {e-mail:alimohmd@theory.ipm.ac.ir},
A. Shariati$ ^{b,c}$}
\vskip 1cm
{\it $^a$ Department of Physics, University of Tehran, North Karegar,} \\
{\it Tehran, Iran }\\
{\it $^b$ Institute for Advanced Studies in Basic Physics , P.O.Box 159 ,}\\
{\it  Gava Zang , Zanjan 45195 , Iran }\\
{\it $^c$ Institute for Studies in Theoretical Physics and Mathematics,}\\
{\it P.O.Box 19395-5531, Tehran, Iran}\\
\end{center}

\vskip 2cm
\begin{abstract}
\noindent
In the context of the Einstein--Cartan--Dirac model, where the torsion of the
space--time couples to the axial currents of the fermions, we study the effects
of this quantum--gravitational interaction on a massless neutrino beam crossing
through a medium with high number density of fermions at rest. We calculate the
reflection amplitude and show that a specific fraction of the
incident neutrinos reflects from this potential if the polarization of the medium
is different from zero. We also discuss the order of magnitude of the fermionic
number density in which this phenomenon is observable,
in other theoretical contexts, for example the
strong--gravity regime and the effective field theory approach.

\end{abstract}

\end{titlepage}
\newpage
\section{Introduction}
Studying the physics of gravity with torsion and especially the interaction of
torsion with a spinor field, has attracted attention for a long time [1-5].
Recently the interest in these theories has increased because of formal
development of string theory. The low--energy limit of string theory has a
an antisymmetric tensor field of the third rank which is usually associated
with torsion of space--time \c{gr}.

One of the important examples of torsion--fermion interaction, is the interaction
of torsion of space--time with neutrinos. These kinds of investigations go back
also to several years ago. For example in ref. \c{sab1} (see also \c{sab2}),
the effect of torsion on neutrino oscillation has been studied by assuming
that the torsion eigenstates, i.e., the eigenstate of the interaction part of
the Hamiltonian , are different from the weak interaction eigenstates.
Recently, the contribution of the torsion of space--time on standard
neutrino oscillation has been studied in the context of Einstein--Cartan--Dirac
theory and for
the case where the mass eigenstates are different from the weak interaction
eigenstates (which is assumed to be the same as the torsion eigenstates).
The situations in which the torsion effect on neutrino oscillation
is as important as the neutrino mass effect have also been discussed \c{ali}.

In this paper, we want to study the effect of torsion--neutrino interaction
potential on the propagation of a massless neutrino beam crossing through a
region (object) which has torsion. As we will see, the axial--currents of
all kinds of fermions, including the neutrinos itself, couple to the totally
antisymmetric part of contorsion, and this coupling produce a potential barrier
with width equal to the radius of that object, if the average spin component
of the fermions, in some direction, is different from zero.
We will show that this kind of
potential, affects the incident massless neutrino flux and reflect a specific
fraction of it. Note that this reflection is not trivial, because if one considers
a constant potential barrier in the Dirac problem of a massless particle, it
can be shown that this barrier can not affect the incident beam and therefore
a constant potential barrier is transparent
for massless particles. We think that this is an interesting phenomenon, as
the only known interaction which can affect an incident massless neutrino beam
is the standard weak interaction, and now we see that this completely gravitational
effect can also produce a quantum mechanical reflection. On the other hand, this
effect predicts the lack of neutrino flux when crossing through
a region which has a high density polarized fermions.

The plan of the paper is as follows. In section 2, we briefly discuss the
Einstein--Cartan--Dirac theory and derive the above mentioned interaction term,
and in section 3, we solve the simple problem of crossing a massless neutrino
beam through this gravitational potential barrier. At the end, we discuss the
order of magnitude of this effect in different theoretical frameworks.

\section{Einstein-Cartan-Dirac theory}

Consider a four dimensional manifold $U^4$ which is specified by two
independent tensor fields,
the Riemannian metric $g_{\m\n}$ and the connection $\G^\m_{\a\b}$
where its most general form, compatible with $g_{\m\n}$, is
\be \la{1}
\G^\a_{\m\n} = \Chr{\a}{\m\n} + K^\a_{\m\n},
\ee
where
\be \la{2}
\Chr{\a}{\m\n} = \half g^{\a\b} \left( -g_{\m\n,\b} + g_{\b\m,\n}
+ g_{\n\b,\m} \right),
\ee
is the usual Christoffel symbol, and $K^\a_{\m\n}$ is the contorsion of $U^4$.
The contorsion tensor is related to the torsion tensor as follows
\be \la{3}
K^\a_{\m\n} = \half g^{\a\b} \left( T_{\b\m\n} + T_{\m\b\n}
+ T_{\n\b\m} \right),
\ee
and the torsion itself is the antisymmetric part of the connection
\be \la{4}
T^\a_{\m\n} = \G^\a_{\m\n} - \G^\a_{\n\m}.
\ee
In this way, it is obvious that the differential geometry of $U^4$  is determined
by two independent tensors $g_{\m\n}$ and $K^\a_{\m\n}$ (or equivalently
$T^\a_{\m\n}$).

If one decomposes the contorsion $K_{\a\m\n}$ as following
\be \la{5}
K_{\a\m\n} = \frac{1}{3} \left( g_{\a\m}\tau_\n - g_{\n\m}\tau_\a \right)
+ \frac{1}{2} \A^\sigma \E_{\sigma\a\m\n} + U_{\a\m\n},
\ee
it can be shown that the independent vectors $\tau_\m$ and $A^\sigma$ satisfy
\be \la{6}
\tau_\m = g^{\a\b} K_{\a\b\m},
\ee
\be \la{7}
\A^\sigma = \frac{1}{3} \E^{\sigma\a\m\n}
K_{\a\m\n},
\ee
and $U_{\a\m\n}$ has the following properties
\be \la{8}
U_{\a\m\n} = -U_{\n\m\a}, \;\; g^{\a\m}U_{\a\m\n} = 0,
\;\; \E^{\sigma\a\m\n}U_{\a\m\n} =0.
\ee
In the above equation, $\E_{\sigma\a\m\n}$ is the totally antisymmetric
pseudo-tensor of rank 4. Now if we calculate the scalar curvature of $U^4$,
using (\r{1}) and (\r{5}-\r{8}), obtain
\be \la{9}
R = ^0R - \frac{2}{\sqrt{g}} \partial_\k(\sqrt{g}\tau^\k)-
\left(\frac{2}{3} \tau^2 + \frac{3}{2}\A^2 + U_{\a\m\n}U^{\m\n\a} \right),
\ee
where $^0R$ is the scalar curvature of the same manifold but with vanishing
torsion, i.e., ${\G^{(0)}}^\a_{\m\n}=\Chr{\a}{\m\n}$, and
$\sqrt{g}=[-{\rm det }(g_{\m\n})]^{1/2}$.

If we couple a spin--1/2 particle to this space--time, the resulting theory
is known as the Einstein-Cartan-Dirac (ECD) theory, and its action is defined by
\be \la{10}
I_{{\rm ECD}}=
- \frac{c^3}{16\pi G}\int d^4 x \sqrt{g} R + \sum_j\int d^4 x \sqrt{g} (-\hbar)
 \bar{\psi_j}\left(e_a^\m \g^a (\partial_\m + \Gamma_\m )
+ i\frac{m_j c}{\hbar} \right) \psi_j ,
\ee
where $a$ is tetrad index and $\G_\m$ is the spin connection
\be \la{11}
\G_\m =-\frac{i}{8} [\g^a, \g^b] e_{a}^{\n} e_{b\n;\m},
\ee
and the sum is over different types of fermions.
In the above equation, $;\m$ denotes the covariant derivative on $U^4$
\be\la{12}
e_{b\n;\m} := e_{b\n,\m} - \G^\l_{\n\m} e_{b\l}
= e_{b\n,\m} - \Chr{\l}{\n\m} e_{b\l} - K^\l_{\n\m} e_{b\l}.
\ee
It can be easily shown that the variation of the action (\r{10}) with respect
to $\bar{\psi_j}, A^\m , \tau^\m$, and $U_{\a\m\n}$ leads to the following
equations of motion, respectively, [8,9]
\be\la{13}
\g^\m\partial_\m\psi_j+i\frac{m_j c}{\hbar}\psi_j +\frac{i}{4}\g_5A_\m\g^\m
\psi_j=0,
\ee
\be\la{14}
A^\m=\frac{12\pi\hbar G}{c^3}\sum_j(J_j)^\m_5,
\ee
\be\la{15}
\tau^\m =0,
\ee
\be\la{16}
U_{\a\m\n}=0,
\ee
where $(J_j)^\m_5=\bar{\psi_j}\g^\m\g_5\psi_j$ is the axial-- current of
the $j$--th Dirac field.\\
Therefore, in the context of ECD theory:\\
1) the axial--currents of the fermions of a region are the source of the torsion
of the space--time of that region,\\
2) when a beam of neutrinos crossing through this region, interacts with the
fermionic matter via ${\cal L}_{{\rm int}}=\frac{i}{4}A_\m\bar
{\psi}_{\rm neutr.}
\g_5\g^\m\psi_{\rm neutr.}$, {\it even the neutrinos are massless}.

\section{Massless neutrino reflection}

In order to study the effect of ${\cal L}_{{\rm int}}$ on a neutrino beam, let
us consider the simple case where the massless neutrino beam crosses through
a spherical object with radius $R$ and with fermionic matter all {\it in at rest}.
So we must first calculate the vector $A^\m$ of this medium. In chiral
representation, the spinor wavefunction of a particle of mass $m$ and zero
momentum, ${\bf p}=0$, is
$\psi =\sqrt{\rho}\left( \begin{array}{c}\chi\\ \chi\end{array} \right)
e^{-iEt/\hbar}$, where $\rho$ is the number density of the particles and $\chi$
is a two--component spinor that must be chosen. We assume that the spin of the
particles is aligned along the unit vector $\hat{{\bf s}}$ which is characterized
by polar and azimuthal angles $\beta$ and $\a$, respectively. We choose $\chi$
to be the eigenspinor of ${\bf\sigma}.\hat{{\bf s}}$, i.e.,
 ${\bf\sigma}.\hat{{\bf s}}\chi =\chi$, so $\chi =\left( \begin{array}{c}
{\rm cos}(\beta /2)e^{-i\a /2}\\{\rm sin}(\beta /2)e^{i\a /2}\end{array}\right)$.
In this way $J^\m_5=\bar{\psi}\g^\m\g_5\psi$ becomes $(0,-2\rho \hat{{\bf s}})$,
and therefore
\be\la{17}
A^\m =(0,-\frac{24\pi\hbar\rho G}{c^3}\hat{{\bf s}}).
\ee
If there are more than one kind of fermion in the medium, we must also add
their contributions to the above relation. In deriving (\r {17}), we assume that
all the fermions are polarized along the same direction $\hat{{\bf s}}$. If the
situation is not so, we must put the average value of spins in (\r {17}). Therefore
${\cal L}_{{\rm int}}$ vanishes for completely random distribution of spins,
i.e., ${\cal L}_{{\rm int}}=0$ {\it for unpolarized medium}.

Now working in chiral representation, the equation of motion (\r {13}) leads to
the following Hamiltonian equation for the wavefunction of massless ($m_\n =0$)
neutrinos
\be\la{18}
(H_0+V_T)u({\bf p})=Eu({\bf p}),
\ee
where $u({\bf p})$ is defined in $\psi ({\bf r},t)=e^{-i/\hbar (Et-{\bf p}.
{\bf r})}u({\bf p})$. Also
\be\la{19}
H_0=\left( \begin{array}{cc}-c{\bf \sigma .p}&0\\ 0&c{\bf\sigma .p}\end{array}
\right) ,
\ee
and
\be\la{20}
V_T=-\frac {\hbar c}{4}
\left( \begin{array}{cc}{\bf \sigma .A}&0\\ 0&{\bf\sigma .A}\end{array}\right)
=K\left( \begin{array}{cc}{\bf \sigma .{\hat s}}&0\\ 0&{\bf\sigma .{\hat s}}
\end{array}\right) ,
\ee
where we have used the expression (\r {17}) for $A^\m$, and $K$ is the coupling
constant of this model (ECD model) and is equal to
\be\la{21}
K_{{\rm ECD}}=\frac {6\pi\rho G\hbar^2}{c^2}.
\ee
Note that in eq.(\r {20}), ${\bf \sigma}$ is the spin vector of the incoming
neutrinos, while $\hat{{\bf s}}$ is the polarization direction of the target
fermions. For simplicity, we take the momentum direction of neutrinos as
$z$--direction and the polarization direction of target particles as $x$--direction.
It can be shown that the $z$--component of $\hat{{\bf s}}$ does not contribute
in neutrino scattering, so if the neutrinos and the target fermions have the same
polarization direction, the fermionic medium becomes {\it transparent} for
the incident massless neutrino beam.

In this way we find the following solution for different regions:\\
for $z\leq 0$ region
\be\la{22}
\psi_1=e^{ipz}\left( \begin{array}{c}0\\ 1\\ 0 \\ 0 \end{array}\right)
+Ae^{-ipz}\left( \begin{array}{c}1\\ 0\\ 0 \\ 0 \end{array}\right)
+Be^{-ipz}\left( \begin{array}{c}0\\ 0\\ 0 \\ 1 \end{array}\right),
\ee
for $0\leq z\leq R$ region
\be\la{23}
\psi_2=Ce^{iqz}\left( \begin{array}{c}\a \\ 1\\ 0 \\ 0 \end{array}\right)
+De^{iqz}\left( \begin{array}{c}0\\ 0\\ 1 \\ \a \end{array}\right)
+Ee^{-iqz}\left( \begin{array}{c}1\\ \beta \\ 0 \\ 0 \end{array}\right)
+Fe^{-iqz}\left( \begin{array}{c}0\\ 0\\ \beta \\ 1 \end{array}\right),
\ee
and for $z\geq R$ region
\be\la{24}
\psi_3=Ge^{ipz}\left( \begin{array}{c}0\\ 1\\ 0 \\ 0 \end{array}\right)
+He^{ipz}\left( \begin{array}{c}0\\ 0\\ 1 \\ 0 \end{array}\right).
\ee
In the above equations $A,\cdots ,H$ are coefficients that must be determined,
$E=pc$ is the energy of the incident neutrino, $q$ is defined in
\be\la{25}
E=\sqrt{K^2+c^2q^2},
\ee
and $\a$ and $\beta$ are
$$
\a =\frac{E-cq}{K},
$$
\be\la{26}
\beta =\frac{K}{E+cq}.
\ee
Writing down the continuities of the solutions in $z=0$ and $z=R$, determine the
coefficients as following
\be\la{27}
B=D=F=H=0,
\ee
so the neutrinos have {\it no spin--flip}, and
\bea\la{28}
&C&=\frac{1}{1-\a\beta e^{2iqR/\hbar}},\cr\cr
&A&=\a C(1-e^{2iqR/\hbar}),\cr\cr
&E&=-\a Ce^{2iqR/\hbar},\cr\cr
&G&=(1-\a\beta )Ce^{i(q-p)R/\hbar}.
\eea
Note that the result (\r {27}) is a consequence of the fact that a massless
particle has definite chirality even in the presence of torsion, in other
words, the Hamiltonian $H_0+V_T$ commutes with $\g_5$.
Therefore the probability of transmission of neutrino beam is (using (\r {25})
and (\r {26}))
\bea\la{29}
|G|^2&=&\frac{1}{1+{1\over 2}\left({K\over cq}\right)^2(1-{\rm sin}2qR/\hbar )}
\cr\cr &=&1-{1\over 2}\left({K\over E}\right)^2(1-{\rm sin}2qR/\hbar )+\cdots ,
\eea
where in the last step we assume that $E=pc\gg K$. In this way we find that the
specific fraction of the incident neutrinos reflect due to this quantum--gravitational
effect, and we have a flux reduction when a massless neutrino beam crosses through
a region with non--zero spin polarization.

Now let us discuss the order of magnitude of this effect in different theoretical
frameworks.\\
{\bf 1.} In ECD theory, in which we have worked until now, $K$ is defined in
eq.(\r {21}) and is equal to
\be\la{30}
K_{{\rm ECD}}({\rm ev})=10^{-69}\rho ({\rm cm}^{-3}).
\ee
So for neutrinos with energy $E_\n \sim O(1)$ev, this effect becomes significance
only when $\rho \sim 10^{69}$ cm$^{-3}$ (see eq. (\r {29})).\\
{\bf 2.} The torsionic contact interaction Lagrangian between two spin half
particles is formally identical to the weak interaction Lagrangian and may be
written in the $(V-A)$ form, if at least one of the two fermions is massless.
This suggest that the spin torsion coupling constant $G_T$, be also identified
with the weak interaction Fermi constant, i.e., $G\rightarrow G_T\approx 10^{31}G$
[7,10-12]. Therefore
\be\la{31}
K_{{\rm V-A}}({\rm ev})=10^{-38}\rho ({\rm cm}^{-3}).
\ee
So in this context, this effect is observable for matters with
$\rho \sim 10^{38}$ cm$^{-3}$.\\
{\bf 3.} Inside the collapsed matter \c{sab3} and in the early stage of the
universe \c{sab4}, the gravity is in the strong regime, in which
$G\rightarrow G_{{\rm SG}}\approx 10^{39}G$ \c{siv}. In these cases
\be\la{32}
K_{{\rm SG}}({\rm ev})=10^{-30}\rho ({\rm cm}^{-3}).
\ee
and the torsion--neutrino interaction becomes important for
$\rho \sim 10^{30}$ cm$^{-3}$.\\
{\bf 4.} Another approach in studying the interaction of torsion and fermions, is
the effective field theory (EFT) approach which have been discussed in \c{bel}.
In this approach, the simplest action which includes all possible terms satisfying
the symmetries of the theory has been considered. However, as far as one is
interested in the low energy effect, the high derivative vertices are suppressed
by the huge massive parameter which should be introduced for this purpose. In the
heavy torsion limit, it can be shown that the Lagrangian of this model has the
following contact four--fermion interaction term
\be\la{33}
{\cal L}_{{\rm int}}=-\frac{\eta^2\hbar^3}{2M_{{\rm ts}}^2c}
(\bar{\psi }\g_5\g^\m\psi)(\bar{\psi }\g_5\g_\m\psi),
\ee
in which the parameter $\eta/(M_{{\rm ts}}c^2)$ has the following limit \c{bel}
\be\la{34}
\frac{\eta}{M_{{\rm ts}}c^2}\leq 1 \ \ {\rm Tev}^{-1}
\ee
Now if we substitute the expression (\r {14}) for $A^\m$ in eq.(\r {13}), it is
clear that we can recover the heavy torsion limit of EFT approach, if we replace
$12\pi\hbar G/c^3$ by $(2\eta\hbar /(M_{{\rm ts}}c))^2$. In this way, the coupling
constant of our model becomes
\be\la{35}
K_{{\rm EFT}}=2\left( \frac{\eta}{M_{{\rm ts}}c^2}\right)^2(\hbar c)^3\rho .
\ee
Note that this replacement is in fact the replacement of the Planck scale energy
($10^{16}$ Tev) with $1$ Tev. Using $M_{{\rm ts}}c^2/\eta \simeq 1$Tev, it is
found that
\be\la{36}
K_{{\rm EFT}}({\rm ev})=10^{-38}\rho ({\rm cm}^{-3}),
\ee
and therefore in the context of EFT, the reflection of the neutrino beam is
observable if $\rho \sim 10^{38}$ cm$^{-3}$.

\vspace {1cm}

\noindent{\bf Acknowledgement}

\noindent M. Alimohammadi would like to thank the research council of the
University of Tehran, for partial financial support.

\end{document}